\documentclass[prb,twocolumn,showpacs,amsmath,amssymb,superscriptaddress]
{revtex4}

\usepackage[dvips]{color}
\usepackage{graphicx}

\begin{document}

\title{Time-of-flight observables and the formation of \\
Mott domains of fermions and bosons on optical lattices}
  
\author{M. Rigol}
\affiliation{Physics Department, University of California, Davis,
CA 95616, USA}
\author{R. T. Scalettar}
\affiliation{Physics Department, University of California, Davis,
CA 95616, USA}
\author{P. Sengupta}
\affiliation{Department of Physics, University of Sothern California, 
Los Angeles, CA 90089, USA}
\author{G. G. Batrouni}
\affiliation{Institut Non-Lin\'eaire de Nice, UMR 6618 CNRS,
Universit\'e de Nice--Sophia Antipolis, 1361 route des Lucioles,
06560 Valbonne, France}

\begin{abstract}
We study, using quantum Monte Carlo simulations, the energetics of the
formation of Mott domains of fermions and bosons trapped on
one-dimensional lattices. We show that, in both cases, the sum of
kinetic and interaction energies exhibits minima when Mott domains
appear in the trap. In addition, we examine the derivatives of the
kinetic and interaction energies, and of their sum, which display
clear signatures of the Mott transition. We discuss the relevance of
these findings to time-of-flight experiments that could allow the
detection of the metal--Mott-insulator transition in confined fermions
on optical lattices, and support established results on the 
superfluid--Mott-insulator transition in confined bosons 
on optical lattices.
\end{abstract}


\maketitle

Following the successful loading of ultracold Bose gases on optical
lattices,\cite{greiner02,stoferle04} a similar scenario has been
achieved in recent experiments with ultracold 
fermions.\cite{modugno03,ott04,kohl05,moritz05} 
In Refs.~\onlinecite{modugno03} and \onlinecite{ott04}
an optical lattice was superposed on highly
anisotropic traps to study transport properties and localization
effects in ideal Fermi gases and Bose--Fermi mixtures.  More recently,
Hubbard- and Luttinger-type systems have been realized by loading
fermionic atoms on three-~\cite{kohl05} and two-dimensional~\cite{moritz05} 
optical lattices, respectively. Now that such experimental
setups are available, one of the most prominent goals is the
achievement of the metal-Mott-insulator transition~\cite{mott37} in
confined fermions on optical lattices.\cite{rigol03,rigol04_1}

Ultracold fermions on optical lattices are an almost ideal realization
of the Hubbard model, in which all parameters can be controlled with
very high precision. In this work, we are interested in the
one-dimensional (1D) regime,\cite{stoferle04,moritz05} where the
Hamiltonian can be written as
\begin{eqnarray}
H &=& -t \sum_{i,\sigma} \left( c^\dagger_{i\sigma} c^{}_{i+1
\sigma} + c^\dagger_{i+1\sigma} c^{}_{i\sigma} \right) 
+ U \sum_i n_{i \uparrow} n_{i \downarrow}
\nonumber \\ &&+ V_2 \sum_{i} x_i^2\ n_{i}, 
\label{HubbF}
\end{eqnarray}
where $c^\dagger_{i\sigma}$, $c_{i\sigma}$ are the creation and
annihilation operators of a fermion in a pseudospin 
state $\sigma$ at site
$i$ (and position $x_i$), and $n_{i}=\sum_{\sigma} n_{i\sigma}$ 
($n_{i \sigma} = c^\dagger_{i\sigma} c^{}_{i\sigma}$) 
is the particle number operator. The on-site
interaction parameter is denoted by $U$ ($U>0$), the hopping amplitude
by $t$, and $V_2$ is the curvature of the harmonic confining
potential.  In experiments, $t$ and $U$ can be modified by changing the
intensity of the laser beams that produce the lattice. In addition,
$U$ can be also controlled separately by means of a Feshbach 
resonance,\cite{kohl05,moritz05} which is the mechanism we consider below. 
Our quantum Monte Carlo (QMC) simulations of the fermionic Hubbard 
model were performed using a projector algorithm~\cite{sugiyama86,sorella89} 
along the lines described in Refs.~\onlinecite{rigol03} and 
\onlinecite{rigol04_1}.

In the absence of a confining potential ($V_2=0$), the phase diagram
of this model, Eq.\ (\ref{HubbF}), for $U>0$, consists of a
Mott-insulating (MI) phase at half-filling ($n=1$), trivial band
insulating phases for $n=0,2$, and a metallic phase for any other
density. In the presence of a confining potential $V_2\neq0$, these
phases coexist in space separated regions.\cite{rigol03,rigol04_1} 
Local measurements may then provide an
experimental proof of the existence of MI domains in the 
trap.\cite{rigol03,rigol04_1} A recent proposal~\cite{liu05} also includes
collective oscillations as a possible way to detect their
formation. However, the demonstration of the MI for fermions is more 
difficult than in the bosonic case. This is because the appearance of
Mott regions for bosons can be detected experimentally by means of
standard time-of-flight (TOF) measurements,\cite{greiner02,stoferle04}
while in the fermionic case the momentum
distribution does not exhibit any distinguishing feature when MI
plateaus develop in the trap.\cite{rigol04_1}

We propose here a novel way to detect the metal-MI transition 
in fermionic systems. We find that the sum of kinetic and interaction 
energies, which can be measured in TOF experiments, and their 
numerical derivatives show distinctive signatures of the formation 
of Mott domains of fermions as well as bosons trapped on optical 
lattices. Hence, energy measurements provide a unified way to deal 
with the Mott transition in confined atoms (fermions or bosons) 
on optical lattices.

Bosons on 1D optical lattices can be described by the boson Hubbard
model~\cite{Jaksch98}
\begin{eqnarray}
\nonumber
H &=&-t\sum_{i} \left(a^{\dagger}_i a_{i+1} +
a^{\dagger}_{i+1}a^{}_i \right) 
+ \dfrac{U}{2}\sum_i n_i(n_i-1)
\\ & & + V_2\sum_i x_i^2\ n_i,
\label{HubbB}
\end{eqnarray}
where $a^\dagger_{i}$, $a_{i}$ are the creation and annihilation
operators of a bosonic atom at site $i$, and $n_{i} = a^\dagger_{i}
a_{i}$ is the particle number operator. The hopping parameter, on-site
interaction, and curvature of the confining potential are denoted as
in the fermionic case. The QMC simulations of the boson Hubbard model
were done using the stochastic series expansion (SSE) method~\cite{sse} 
in the grand canonical ensemble in which an additional chemical
potential term in Eq.\ (\ref{HubbB}) is used to control the filling.

\begin{figure}[h]
\begin{center}
\includegraphics[width=0.45\textwidth]
{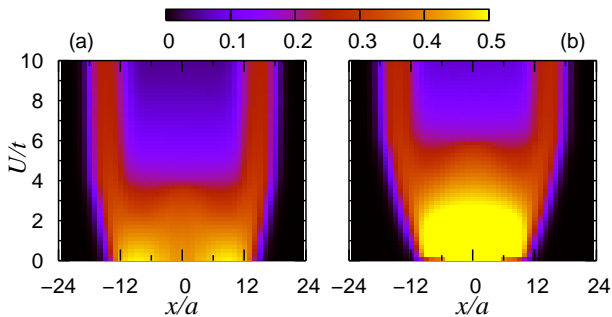}
\end{center} \vspace{-0.6cm}
\caption{(color online). Intensity plot of the density fluctuations
$\Delta_i=\langle n_i^2\rangle-\langle n_i\rangle^2$ vs position and
interaction strength, in fermionic (a) and bosonic (b) systems with 30
atoms on 48 lattice sites and $V_2a^2=0.015t$ ($a$ is the lattice constant).  
The lattice
size was chosen large enough so that the particle density is zero
at the boundaries.  For large
$U$, the constant small value of $\Delta_i$ in the center of the trap
signals the formation of the MI core. In (b), for low values of $U/t$,
we have truncated $\Delta_i$ at 0.5, i.e., for low on-site repulsive
interactions the bosonic fluctuations of the density exceed 0.5.  }
\label{perfil3D}
\end{figure}

In Fig.\ \ref{perfil3D}(a), we show the evolution of the variance of
the density ($\Delta_i=\langle n_i^2\rangle -\langle n_i\rangle^2$) in
a 1D system of fermions when the on-site repulsion is increased using
a Feshbach resonance.  At $U=0$, $n_i<2$ throughout the trap so that
no local ``band insulator'' is present. This can be easily checked
experimentally measuring the momentum distribution function, which
should exhibit a Fermi momentum $k_F$, i.e, $n_{k>k_F}=0$.\cite{rigol04_2} 
(The band insulator always produces a finite
occupation of all momentum states within the Brillouin 
zone.)\cite{rigol04_2} For $U>4t$, a region of small $\Delta_i$,
indicating suppressed (but finite) quantum fluctuations of the density,
appears in the center of the trap showing the establishment of a MI
there.\cite{rigol03,rigol04_1}  Notice that in a trap, in contrast 
to periodic systems at half-filling, the MI appears only for a finite 
value of $U/t$.

The corresponding evolution of $\Delta_i$, for a bosonic system with
the same trap parameters and number of particles as the fermionic
case, Fig.\ \ref{perfil3D}(a), is depicted in Fig.~\ref{perfil3D}(b).
$\Delta_i$ again signals the formation of a MI plateau~\cite{batrouni02} 
when $U\sim 6.1t$, i.e., a value that is $\sim 2t$
larger than in the fermionic case and also larger than the critical
value $U/t=3.5$ in the periodic case.\cite{kuhner98} Although for small 
values of $U$, quantum fluctuations of the density for bosons are much 
larger than for fermions (in the figure we have truncated them at 
$\Delta=0.5$), when $U$ is increased both systems behave analogously, 
revealing the similarities between bosons and fermions in 1D.

Several approaches based on TOF measurements have been followed in
experiments to detect the formation of Mott domains of bosons.  
They include the disappearance of the interference pattern,\cite{greiner02} 
the increase of the full width at half 
maximum,\cite{stoferle04,kollath04,wessel04} and more recently the behavior
of the visibility.\cite{gerbier05,sengupta05} They are all related 
to the momentum distribution function $n_k$ of the trapped system,
which is measured in TOF experiments. We have studied these observables 
in the fermionic case, and found no signature of the formation of MI 
domains.

A further quantity related to $n_k$ that can be also obtained
experimentally is the kinetic energy,
\begin{equation}
E_K=-2t \sum_{k } n_k \cos(ka),
\end{equation}
where $a$ is the lattice parameter, and $-2t \cos(ka)$ is the lattice
dispersion relation.

\begin{figure}[h]
\begin{center}
\includegraphics[width=0.48\textwidth]
{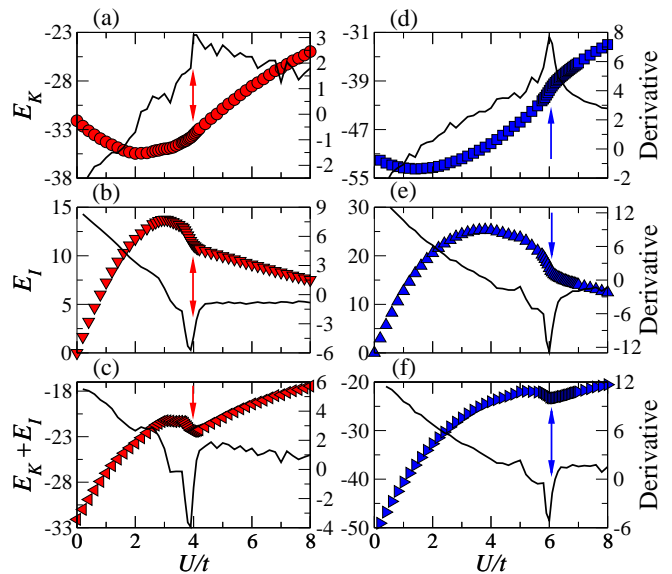}
\end{center} \vspace{-0.6cm}
\caption{(color online). (a), (d) Kinetic energy ($E_K$), (b), (e)
interaction energy ($E_I$), and (c), (f) the sum $E_K+E_I$, for
fermions (a-c) and bosons (d-f) in the systems of Fig.\
\ref{perfil3D} (30 atoms and $V_2a^2=0.015t$). 
Energies are given in units of $t$. Continuous lines depict 
numerical derivatives of the energies. The arrows signal the emergence
of a MI core. The QMC errors are much smaller than our symbol sizes 
throughout this work.}
\label{fig2}
\end{figure}

The evolution of the kinetic energy with $U$, for the fermionic system
in Fig.\ \ref{perfil3D}, is shown in Fig.\ \ref{fig2}(a). $E_K$ does not 
exhibit any special feature at the point where the MI appears in the 
system. In contrast, the interaction energy ($E_I$), depicted in Fig.\
\ref{fig2}(b), displays a sharp drop just before the Mott plateau
sets in. After that, the decrease of $E_I$ occurs slowly, even
more slowly than the reduction of the modulus of the kinetic
energy. This suggests that if one turns the trap off and lets the
particles expand on the lattice, so that the interaction energy is
converted to kinetic energy, the resulting kinetic energy
$E'_K=E_K+E_I$ can be used to determine when the MI forms.  
Figure \ref{fig2}(c) shows that $E'_K$ exhibits
a local minimum where the Mott core appears in the system.

An enhanced signal of the features observed in $E_I$ and $E_I+E_K$ can
be obtained by calculating their numerical derivatives. As shown in
Figs. \ref{fig2}(b) and \ref{fig2}(c), a large reduction in the
derivatives of both quantities occurs just before the MI sets in the
center of the trap. The minimum in $E_I+E_K$ is signaled by a vanishing
derivative. The derivative of the kinetic energy, 
Fig. \ref{fig2}(a), exhibits another signal, a weak maximum and a steady 
decrease after the formation of the MI core.

Using a Feshbach resonance, one can also obtain a final $E'_K$
exhibiting the features observed in $E_I$, Fig.\ \ref{fig2}(b).  To
that end, just before turning off the trap, one could increase $U$ (and
hence $E_I$) by a large constant factor, in a timescale much smaller
than the scale set by the hopping parameter ($\hbar/t$). In this case
$E'_K$, measured after the expansion on the lattice, would be
dominated by $E_I$ thus allowing the observation of the interaction
strength at which the MI appears in the trap.\cite{tilman}

\begin{figure}[h]
\begin{center}
\includegraphics[width=0.39\textwidth]
{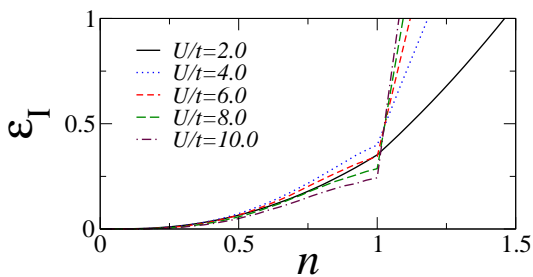}
\end{center} \vspace{-0.7cm}
\caption{(color online). Ground-state on-site interaction 
energy per lattice site ($\epsilon_I$) of fermions as a function 
of the density [obtained from a periodic system (no trap) with 102 
lattice sites], for different values of $U/t$. At $n=1$ fermions 
are in the MI phase, 
and for any other value of $n$ they are in a metallic phase.}
\label{fig3}
\end{figure}

In order for the above proposal to work, one needs a fast conversion
of the interaction energy into kinetic energy. We expect this to 
be the case since the reduction of the density during the expansion 
produces a preponderance of empty and singly occupied states over the 
doubly occupied ones, i.e., on-site interactions get strongly suppressed. 
For example, in the ground state, the dependence of the on-site
interaction energy on the density is shown in Fig.\ \ref{fig3}. 
One can see that below $n=0.25$, the interaction energy is negligible,
independent of the on-site repulsion $U$.

\begin{figure}[h]
\begin{center}
\includegraphics[width=0.48\textwidth]
{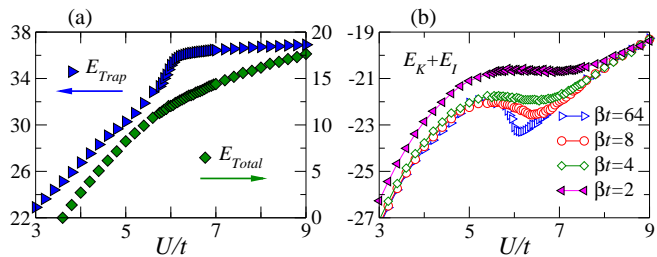}
\end{center} \vspace{-0.7cm}
\caption{(color online). (a) Trap ($E_{Trap}$) and total ($E_{Total}$)
energies corresponding to the systems of Figs.\ \ref{fig2}(d)-\ref{fig2}(f).
(b) $E_K+E_I$ for systems with the same trap parameters of Fig.\ \ref{fig2} 
and different inverse temperatures $\beta=1/k_B T$. $\beta t=64$ corresponds 
to the ground-state results of Figs.\ \ref{fig2}(d)-\ref{fig2}(f).}
\label{fig4}
\end{figure}

Measuring energies to detect the formation of MI domains 
has the advantage of being useful independent of
the statistics of the particles involved.  In Figs.\ 
\ref{fig2}(d)--\ref{fig2}(f),
we also show results corresponding to the bosonic system in Fig.\
\ref{perfil3D}.  The same features discussed above for fermions are 
present in the bosonic case. We should stress here that the minimum 
that signals the formation of the MI in Fig.\ \ref{fig2}(f) is an 
exclusive property of trapped systems, i.e., it is not present in the
periodic case. This minimum is related to the fast increase of the 
trapping energy $E_{\text{Trap}}$ [see Fig.\ \ref{fig4}(a)] produced by the 
formation of Mott domains, which push particles to the outer regions 
of the trap. On the other hand, as expected from the absence of a global 
phase transition,\cite{batrouni02} the total energy of the system 
increases smoothly throughout the formation of the Mott plateau
[Fig.\ \ref{fig4}(a)]. Hence, we expect the above signatures to be 
present also in higher-dimensional systems where most of the 
experiments are carried out.

To what extent will finite temperatures reduce this signal?  We find 
that as long as the temperature is low enough to allow
for the formation of a MI phase, the dips in the energy are present.
In Fig.\ \ref{fig4}(b), we show $E_K+E_I$ for systems 
at different inverse temperatures $\beta=1/k_B T$ (where $k_B$ is the 
Boltzmann constant and $T$ is the temperature). For $\beta t=64$, the system
is essentially in its ground state 
[results in Figs.\ \ref{fig2}(d)--\ref{fig2}(f)]. 
As seen in Fig.\ \ref{fig4}(b), the ground-state minimum of $E_K+E_I$  
is still present (although weakening) with increasing temperature 
($\beta t=8$ and 4) when Mott domains are still observed in the density 
profiles. For high temperatures [$\beta t=2$ in Fig.\ \ref{fig4}(b)], no 
Mott plateau appears in the trap and no minimum is seen in $E_K+E_I$.

\begin{figure}[h]
\begin{center}
\includegraphics[width=0.45\textwidth]
{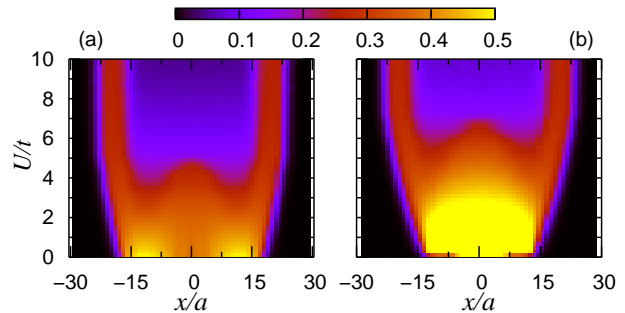}
\end{center} \vspace{-0.6cm}
\caption{(color online). Intensity plot of the density fluctuations 
$\Delta_i=\langle n_i^2\rangle-\langle n_i\rangle^2$ vs position and 
interaction strength in fermionic (a) and bosonic (b) systems with 
40 atoms on 60 latice sites and $V_2a^2=0.01t$. With increasing $U$, 
two Mott domains 
appear at the sides of the central metallic phase before a full Mott
core develops in the center of the trap. In (b), for low values of $U/t$,
$\Delta_i$ has been truncated as in Fig.\ \ref{perfil3D}.}
\label{perfil3D_1}
\end{figure}

\begin{figure}[h]
\begin{center}
\includegraphics[width=0.48\textwidth]
{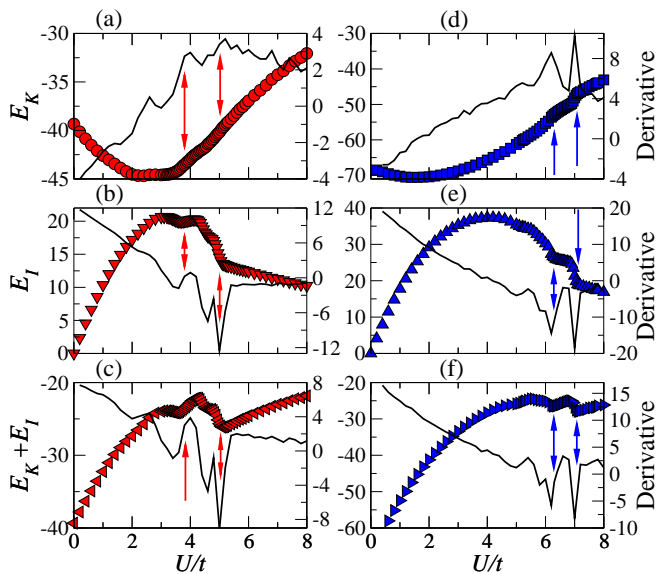}
\end{center} \vspace{-0.6cm}
\caption{(color online). (a), (d) Kinetic energy ($E_K$), (b), (e)
interaction energy ($E_I$), and (c), (f) the sum $E_K+E_I$, for
fermions (a-c) and bosons (d-f) in the systems of Figs.\
\ref{perfil3D_1} (40 atoms and $V_2a^2=0.01t$).  Continuous lines
depict the numerical derivatives of the energies.  The arrows indicate
when two MI domains (lower value of $U/t$) and the MI core (larger
value of $U/t$) emerge in each system.  }
\label{fig6}
\end{figure}

In general, Mott-insulating plateaus can also appear on the sides of 
a central metallic region with $n>1$.\cite{rigol03,rigol04_1,liu05} 
With increasing $U$, these MI regions spread inward and merge
to occupy the core of the 
trap.\cite{rigol03,rigol04_1,batrouni02,kollath04} The evolution of
$\Delta_i$ for a system like this is shown in Fig.\ \ref{perfil3D_1} 
for fermions and bosons with the same trap parameters. The emergence 
of two Mott domains around the central metallic phase is reflected 
by a suppression of $\Delta$ around $U/t\sim 3.8$ for fermions (a), 
and $U/t\sim 6.3$ for bosons (b). The central MI core develops at
$U/t\sim 5.0$ for fermions (a) and $U/t\sim7.1$ for bosons (b).

In Fig.\ \ref{fig6}, we show the kinetic and interaction energies and
their sum for fermions [(a)--(c)] and bosons [(d)--(f)].  Clear local
minima can be seen in $E_K+E_I$, Figs.\ \ref{fig6}(c) and
\ref{fig6}(f), when (i) The two MI domains of fermions (bosons) start to 
develop at the sides of the central metallic (superfluid) phase, 
and (ii) when the Mott core forms in the center of the trap due to 
the merging of the two Mott shoulders.
An increase of $E_K+E_I$ between these two points occurs because 
when the two MI domains surround the central metallic (superfluid) 
region, the transfer of particles to the outer regions costs a finite 
amount of trapping energy. This produces a ``freezing'' of the density 
profiles with increasing $U$ so that $E_K+E_I$ increases by finite amounts 
before particles can be transfered out of the Mott regions to produce a 
full MI in the trap center.\cite{sengupta05} We have also found that the 
behavior of the energies between points (i) and (ii) depends on the trap
parameters. This means that complicated structures may be observed related 
to the discontinuous transfer~\cite{sengupta05} of particles out of the 
central metallic (superfluid) region [see for example the fermionic 
case in Figs.\ \ref{fig6}(b) and \ref{fig6}(c) in contrast to the bosonic case 
in Figs.\ \ref{fig6}(e) and \ref{fig6}(f)]. They are not related to the formation 
of new Mott domains. One can be certain that the full Mott plateau in 
the center of the trap has been formed from the smooth increase of 
$E_K+E_I$ with $U$ after a minimum. (In this case, the density profiles 
almost do not change with $U$.)\cite{rigol04_1}

In Fig.\ \ref{fig6}, we have also shown the numerical derivatives of the 
energies. They also provide clear evidence for the formation of Mott 
domains. In the fermionic case, the derivative of $E_K$ [Fig.\ \ref{fig6}(a)] 
decreases steadily after the full Mott insulator sets in the middle of 
the trap. Two very weak maxima can be also seen at the points signaled 
by the arrows. [They are clearer in the bosonic case, 
Fig.\ \ref{fig6}(b).] The two values of $U/t$ where the derivative of 
$E_K+E_I$, Figs.\ \ref{fig6}(c) and \ref{fig6}(f), increases through 
zero provide a further signature of the two minima in $E_K+E_I$.

In summary, we have shown that minima in $E_K+E_I$ signal the
formation of Mott domains of fermions or bosons in a trap. We have
also studied the signatures that MI phases imprint in the derivatives
of the kinetic and interaction energies, and of their sum. We have
discussed the relevance of our findings to the detection, by means of
TOF measurements, of the formation of Mott domains of fermions and
bosons confined on optical lattices. Our proposal represents a novel
way to detect the metal--Mott-insulator transition using TOF measurements
in fermionic systems, while for bosons the implementation of our 
findings provides an alternative way to support the established 
experimental evidence of the superfluid--Mott-insulator transition.

\noindent
\underbar{Acknowledgments} M.R. thanks T. Esslinger and R. R. P. Singh 
for useful discussions. This work was supported by NSF-DMR-0312261, 
NSF-DMR-0240918, and DE-FG02-05ER46240.

\end{document}